\documentclass[12pt]{iopart}
\usepackage{amsmath}
\usepackage{amssymb}
\usepackage{graphicx}

\usepackage{iopams}  
\begin{document}

\title{Temperature-dependent electron mobility in InAs nanowires}

\author{Nupur Gupta$^{2,3,4,7}$, Yipu Song$^{1,3,4,7}$, Gregory W. Holloway$^{2,3,4}$, Urbasi Sinha$^5$, Chris M. Haapamaki$^6$, Ray R. LaPierre$^6$, Jonathan Baugh$^{1-4}$}

\address{$^1$Department of Chemistry, $^2$Department of Physics and Astronomy, $^3$Institute for Quantum Computing, $^4$ Waterloo Institute for Nanotechnology, University of Waterloo, 200 University Ave. W., Waterloo, ON, Canada, $^5$Raman Research Institute, Sadashivanagar, Bangalore, India, $^6$Department of Engineering Physics, McMaster University, 1280 Main St. W., Hamilton, ON, Canada, $^7$these authors contributed equally to this work. }

\ead{baugh@iqc.ca}

\begin{abstract}
Effective electron mobilities are obtained by transport measurements on InAs nanowire field-effect transistors at temperatures ranging from $10-200$ K. The mobility increases with temperature below $\sim 30-50$ K, and then decreases with temperature above 50 K, consistent with other reports. The magnitude and temperature dependence of the observed mobility can be explained by Coulomb scattering from ionized surface states at typical densities. The behaviour above $50$ K is ascribed to the thermally activated increase in the number of scatterers, although nanoscale confinement also plays a role as higher radial subbands are populated, leading to interband scattering and a shift of the carrier distribution closer to the surface. Scattering rate calculations using finite-element simulations of the nanowire transistor confirm that these mechanisms are able to explain the data. 
\end{abstract}

\maketitle

\section{Introduction}
\indent Semiconductor nanowires grown by the vapour-liquid-solid (VLS) method \cite{dayeh2009,plante2009,schroer2010, joyce2010, dick2010} are the subject of active study, with many potential applications ranging from nanoscale circuits \cite{xiang2006} and gas sensors \cite{du2009} to high-efficiency solar cells \cite{garnett2011,hochbaum2010, lapierre2011a, lapierre2011b}. In particular, InAs nanowires form Ohmic contacts easily\cite{suyatin2007}, and can be grown with low structural defect densities \cite{schroer2010}, giving rise to high electron mobilities \cite{ford2009}, though still low compared to high-quality bulk InAs \cite{rode1971}. The quasi-one-dimensional nature of electron transport at low temperatures \cite{blomers2011} together with a spin-orbit coupling $\sim 40$ times larger than GaAs makes InAs an attractive material for the development of spintronic devices such as electron spin qubits in gate-defined quantum dots \cite{nadjperge2010, baugh2010, schroer2011}. Although transport in InAs nanowires is well-studied \cite{dayeh2010}, the detailed role played by surface states and the surface potential \cite{wieder1974, watkins1995, schrieffer1955, affentauschegg2001} with regard to the electron mobility is not well understood. \\
\indent In this paper, we present electron mobility measurements on low defect density InAs nanowire field-effect transistors (FETs) that show a characteristic temperature dependence. The mobility peaks in the range $3,000-20,000$ cm$^2$V$^{-1}$s$^{-1}$ near $40$ K, with a positive slope at lower temperatures and a negative slope at higher temperatures. Even though acoustic phonon scattering produces a temperature dependence \cite{madelung1964} consistent with the data above $\sim 50$ K, the estimated mobility is much too large (2-3 orders of magnitude) to explain our observations. A similar argument excludes optical phonon scattering as a dominant mechanism in this temperature range (it might dominate at even higher temperatures). We expect this to remain true even in quasi-one-dimensional systems, where phonon scattering is moderately enhanced due to a larger available phase space for scattering \cite{Bruus1993}.  Furthermore, our experimental results are obtained on nanowires with low stacking fault densities, which we confirm using transmission electron microscopy to inspect devices after transport measurements. This excludes stacking faults or twinning defects from explaining the qualitative temperature dependence of mobility. On the other hand, the nanowire geometry suggests that a surface scattering mechanism should be dominant. Surface states are known to be present at densities $\sim 10^{11}-10^{12}$ cm$^{-2}$ eV$^{-1}$ and to act as electron donors. We argue that these positively charged surface states should be more effective at scattering electrons than surface roughness (charge neutral defects), and therefore limit the mobility. Our numerical simulations show that surface charges at the known densities will indeed lead to scattering rates that produce mobilities of the correct order. We find that the decrease in mobility with temperature above $\sim 50$ K can be explained by an increase in the number of ionized surface states due to thermal activation. Consistent with this picture, chemical treatment of the nanowire surface is seen to have a strong effect on the temperature-dependent mobility. Surface roughness scattering, on the other hand, should produce a weaker temperature dependence than what we observe \cite{nag1980}. These results underscore the need for tailored surface passivation techniques \cite{tilburg2010, Haapamaki2012} to reduce the density of surface scatterers and smooth the local electronic potential, leading to increased carrier mobility and more ideal devices for a wide range of quantum transport, nanoscale circuitry and optoelectronics applications. \\
\begin{figure*}[t]
\includegraphics[width= 16cm]{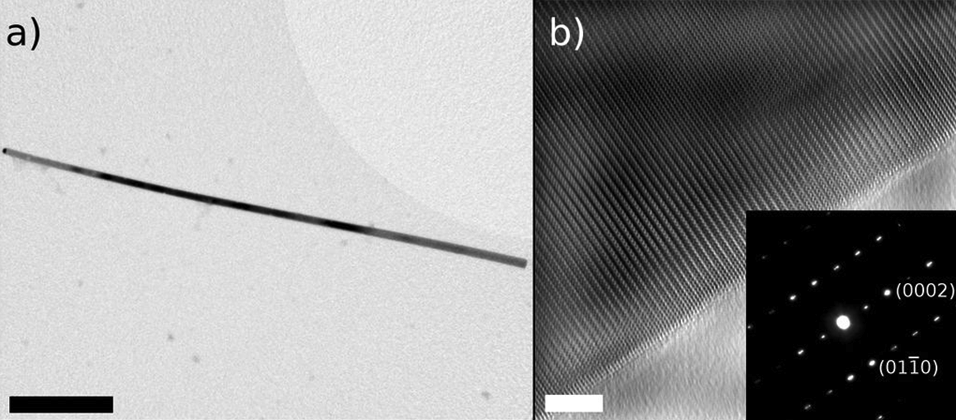} 
\caption{(a) Low and (b) high magnification bright-field TEM images of an InAs nanowire grown by GS-MBE at 0.5 $\mu$m/hr.  Scale bars are 500 nm in (a) and 5 nm in (b). The inset in (b) shows selected area diffraction pattern along the $\left[2\bar{1}\bar{1}0\right]$ zone axis indicating pure wurtzite crystal structure. A majority of wires grown under these conditions had low stacking fault densities $<1~\mu$m$^{-1}$.} \label{fig1}
\end{figure*}
\section{Nanowire growth by gas-source MBE}
\indent InAs nanowires were grown in a gas source molecular beam epitaxy (GS-MBE) system using Au seed particles. A 1 nm Au film is heated to form nanoparticles on a GaAs (111)B substrate. For nanowire growth, In atoms were supplied as monomers from an effusion cell, and As$_2$ dimers were supplied from an AsH$_3$ gas cracker operating at 950$^{\circ}$C. Nanowire growth proceeded at a substrate temperature of 420$^{\circ}$C, an In impingement rate of 0.5 $\mu$m/hr, and a V/III flux ratio of 4. The nanowires grew in random orientations with respect to the GaAs (111)B substrate, possibly due to the large lattice mismatch strain between InAs and GaAs. Transmission electron microscopy (TEM) analysis, shown in \ref{fig1}a, indicated a Au nanoparticle at the end of each nanowire (darker contrast at the left end), consistent with the VLS process.  Most nanowires had a rod-shaped morphology with negligible tapering and a diameter ($\sim20-80$ nm) that was roughly equal to the Au nanoparticle diameter at the top of each nanowire, indicating minimal sidewall deposition. \\
\indent A common occurrence in III-V nanowires is the existence of stacking faults whereby the crystal structure alternates between zincblende and wurtzite, or exhibits twinning, along the nanowire length. Joyce et al. \cite{joyce2010} and Dick et al. \cite{dick2010} have shown that growth parameters in metalorganic chemical vapour deposition (MOCVD) have profound effects on the InAs nanowire crystal phase. Zincblende, wurtzite, or mixed zincblende/wurtzite nanowires were formed by simply tuning the temperature and V/III ratio. We have found that for GS-MBE grown InAs nanowires, stacking faults can be nearly eliminated and pure wurtzite structures can be realized at sufficiently low growth rate $\sim0.5 \mu$m/hr. At higher growth rates, but otherwise identical growth conditions, the InAs nanowires exhibited a much larger fraction of stacking faults on average.  For example, TEM analysis of InAs nanowires grown at a rate of 1 $\mu$m hr$^{-1}$ exhibited an average linear density of stacking faults $\approx 1~\mu$m$^{-1}$.  Similar to GaAs nanowires \cite{fortuna2008,shtrikman2009,joyce2010}, the density of faults diminished dramatically when the growth rate was reduced. Selected area electron diffraction for a typical nanowire (inset of \ref{fig1}b) confirms the pure wurtzite crystal structure and the absence of stacking faults.\\
\begin{figure*}[t]
\includegraphics[width= 16cm]{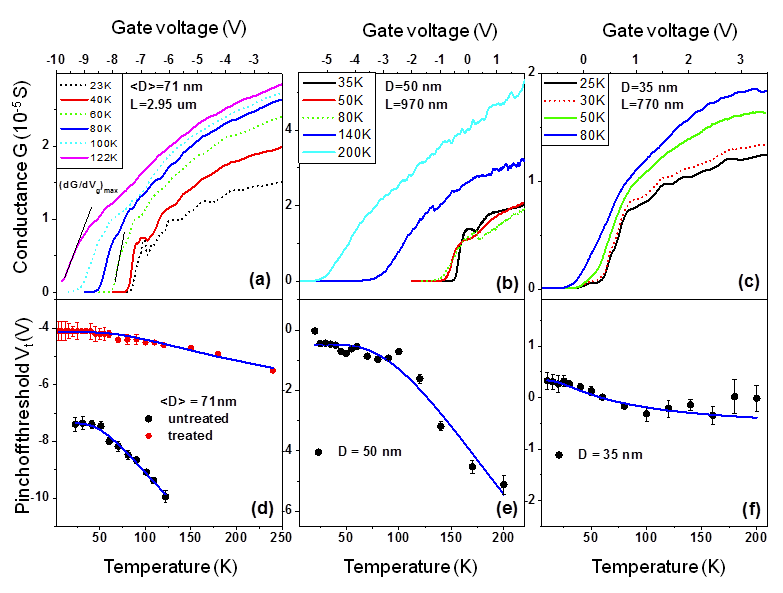} 
\caption{(a-c) Conductance versus backgate voltage for devices $1-3$ at selected temperatures. $D$ is the nanowire diameter and $L$ the FET channel length (device 1 is tapered with an average nanowire diameter $\langle D\rangle=71$ nm). The tangent lines drawn on the $T=122$ K and $T=60$ K traces in (a) indicate the maximum slopes corresponding to peak field-effect mobility. The pinchoff threshold voltage is defined as the intercept between this tangent line and the $G=0$ axis. (d-f) The pinchoff threshold voltages versus temperature extracted from the conductance measurements. In (d), data are shown for device 1 before and after an ammonium sulfide treatment was applied to the FET channel (the data in (a) correspond to the untreated case). The empirical fits in (d-f) are of the form $V_t = V_0 + V_1 e^{-E_a/kT}$, as described in the text.} \label{fig2}
\end{figure*}
\section{Mobility in field-effect transistors}
\indent Field-effect transistors (FETs) were fabricated by mechanically depositing as-grown nanowires on a $175$ nm thick SiO$_2$ layer above a n$^{+}$-Si substrate that functions as a backgate, and writing source/drain contacts for selected wires using electron-beam lithography (schematic device layout is shown in supplementary figure 1). This was followed by an ammonium sulfide etching and chemical passivation process to remove the native oxide and prevent regrowth \cite{suyatin2007} prior to evaporation of Ni/Au contacts. This process yields devices with contact resistance that is small compared to the channel resistance \cite{suyatin2007}. Channel lengths ranged from 0.7 to 3 $\mu$m. Transport measurements were carried out in He vapour in an Oxford continuous flow cryostat from 10 K to room temperature. Bias and gate voltages were applied using a high resolution home-built voltage source, and a DL Instruments current preamplifier was used to measure DC current at a noise floor $\sim 0.5$ pA$/\sqrt{Hz}$. All devices tested at room temperature displayed fully Ohmic I-V characteristics, with resistances typically in the range of $10-200$ k$\Omega$. Gate sweeps were performed at a rate between 3 mV/s (lower temperatures) and 10 mV/s (higher temperatures). Earlier work reported that a sweep rate of 7 mV/s led to very small hysteresis and therefore minimal interface capacitance \cite{dayeh2007b}. Under these conditions, we observe a shift with respect to gate voltage of less than 50 mV upon changing sweep direction, and no observable change in the shape of the conductance curve. Note that FET devices with channel lengths greater than $\sim 200$ nm are known to be in the diffusive transport regime \cite{Zhou2006}.\\
\indent The gate capacitance per unit length was calculated using the expression \cite{wunnicke2006, ford2009, tilburg2010}
\begin{equation}
C'_g = 2\pi \epsilon_0 \epsilon_r/\cosh^{-1}\left(\frac{R+t_{ox}}{R}\right)
\end{equation}\label{eqn3}
where $R$ is the nanowire radius, $\epsilon_0$ is the electric constant, $\epsilon_r=3.9$ is the relative dielectric constant and $t_{ox}$ the thickness of the SiO$_2$ layer, respectively. For the devices studied here, TEM analysis indicated $t_{ox}=175$ nm. The equation above assumes that the nanowire is embedded in SiO$_2$; to compensate for the fact that the nanowire actually sits atop the SiO$_2$ and is surrounded by vacuum ($\epsilon_r=1$), it was shown by Wunnicke \cite{wunnicke2006} that a modified dielectric constant $\epsilon'_r = 2.25$ can be taken. Our numerical simulations, comparing the pinchoff threshold voltages of the FET device calculated with and without SiO$_2$ embedding, confirmed that this is a suitable correction factor. The capacitances based on \ref{eqn3} are listed in Table 1.
\begin{table*}[t!]
\begin{tabular}{c|ccc}
\hline device $\#$ & $D$ (nm)  & ~~ $L$ ($\mu$m)~~ & ~~$C'_g$ (aF$\cdot \mu$m$^{-1}$) ~~\\
\hline\hline
1 &  71  & 2.95 & 50.76 \\
2&  50  & 0.97 & 45.21\\  
3&  35  & 0.77 & 40.52\\     
\hline
\end{tabular}
\caption {Diameters ($D$) and channel lengths ($L$), measured by AFM and TEM, and calculated capacitance per unit length ($C'_g$) for the three main FET devices investigated. Uncertainties in diameter are $\pm 2$ nm (for tapered device 1, $D$ is the average diameter).}  
\label{smtable2}
\end{table*}
\subsection{Results}
\indent We investigated 10 devices to varying levels of detail, and found qualitatively similar results. Here we will focus on three representative devices, denoted 1, 2 and 3 with nanowire diameters $D =$ 71, 50 and 35 nm, respectively. The nanowires in devices 2 and 3 were untapered, whereas the nanowire in device 1 was tapered, with diameter linearly varying from 53 nm to 90 nm across the FET channel (average diameter $\langle D \rangle =$ 71 nm). TEM analysis was carried out on selected devices after transport studies were complete to check for the presence of stacking fault defects. Devices 1 and 3 were found to have zero and one fault, respectively, whereas a fourth device ($D =$ 55 nm) with low mobility was found to have an atypically large fault density (see section~\ref{faults} below). TEM analysis was not performed on device 2. Transport data for an additional high mobility device with $D =$ 50 nm is shown in supplementary figure 3. The channel of device 1 was subjected to an ammonium sulfide etching and passivation treatment, similar to that carried out prior to contacting, after the initial set of transport measurements were completed. Subsequent transport measurements were taken several days later, likely after the native oxide had partially or fully regrown.\\
\indent Figure~\ref{fig2}(a-c) shows conductance $G=I_{sd}/V_{sd}$, where $I_{sd}$ and $V_{sd}$ are the source-drain current and bias, respectively, versus backgate voltage $V_g$ for devices 1, 2 and 3 at selected temperatures. The bias is set to $V_{sd}=1$ mV (similar results are obtained at higher bias). For all three devices, the maximum transconductance $\left(\frac{dI_{sd}}{dV_g}\right)_{max}$ is seen to decrease as temperature is raised above $\sim 30-50$ K. Figures~\ref{fig2}(d-f) show the pinchoff threshold voltages $V_t$ corresponding to the data in figures~\ref{fig2}(a-c), where $V_t$ is defined as the intercept between the maximum slope tangent line and the $G=0$ axis. $V_t$ typically shifts toward more positive gate voltages as temperature decreases, and saturates below $\sim 50$ K. All temperature sweeps reported here were from low to high temperature. We fit the pinchoff threshold data to an empirical function based on thermal activation $V_t = V_0 + V_1 e^{-E_a/kT}$, where $k$ is the Boltzmann constant, typically yielding an $E_a\sim 5-30$ meV. Note that for device 1 in figure~\ref{fig2}(d) we also plot the $V_t$ measured after the chemical treatment was applied to the FET channel. $V_t$ shifted considerably to more positive gate voltage post-treatment, and also showed a weaker temperature dependence. This suggests that the surface potential and density of conduction electrons in the nanowire are controlled in large part by the surface chemistry \cite{du2009}. Post-passivation conductance versus gate curves for device 1 are given in supplementary figure 2. \\
\begin{figure*}[!t]
\includegraphics[width= 16cm]{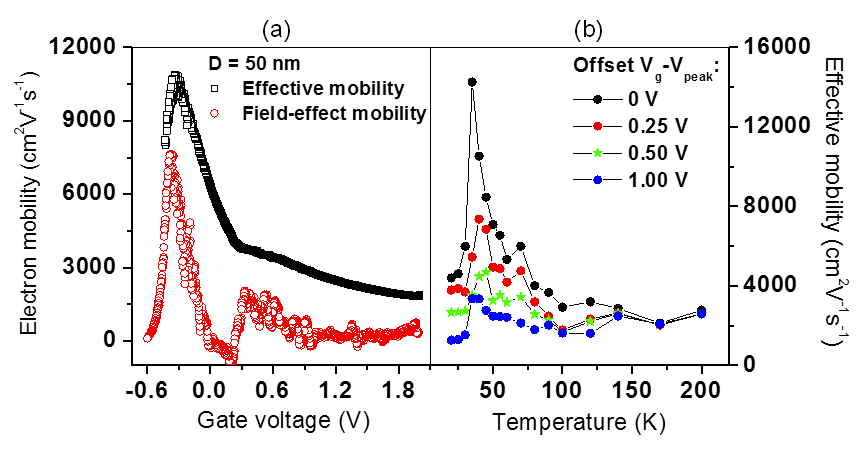} 
\caption{(a) Comparison of the field-effect and effective mobilities for device 2 at $T=40$ K. (b) The temperature dependence of effective mobility for device 2 at different values of gate voltage relative to $V_{peak}$, the gate voltage at which peak mobility occurs. The values at $V_{peak}$ are shown by black dots, at $V_{peak}+0.25$ V by red dots, etc. The mobility at $V_{peak}+0.5$ V (green stars) is near the crossover point between the two slopes seen in the effective mobility in the left panel. } \label{fig3}
\end{figure*}
\begin{figure*}[!t]
\includegraphics[width= 17cm]{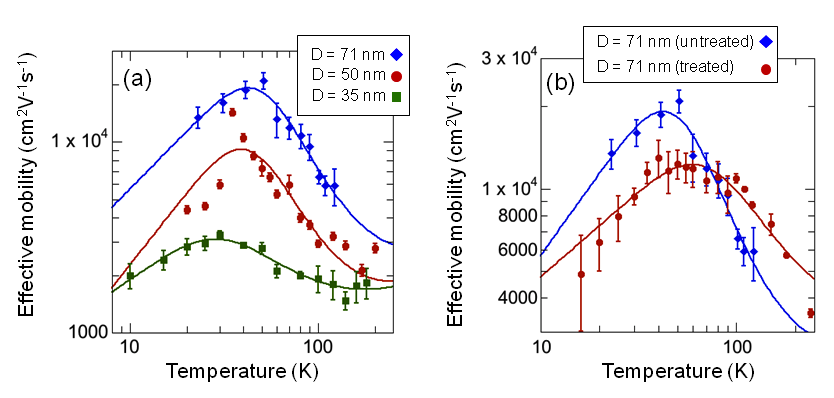} 
\caption{(a) Experimental peak effective mobilities versus temperature for devices $1-3$ (diameters 71, 50, and 35 nm, respectively). The empirical fitting function described in the text (solid lines) is given by $\mu = AT^x (1+Be^{-E_a/kT})^{-2}$, where $x$, $A$, $B$ and $E_a$ are fitting parameters given in the main text. (b) Comparison of peak effective mobilities versus temperature for device 1 before and after an ammonium sulfide etching and passivation treatment was applied to the FET channel. The fitting function is of the same form. For comparison, the pinchoff threshold voltages before and after treatment are shown in figure~\ref{fig2}(d). } \label{fig4}
\end{figure*}
\indent From the measured conductance versus backgate voltage curves, both the field-effect mobility and the effective mobility \cite{ford2009} may be extracted. The field-effect mobility is a lower bound on the effective mobility, and is defined as
\begin{equation}
\mu_{\text{fe}} =q^{-1}\frac{d\sigma}{dn} = \frac{L}{C'_g}\frac{dG}{dV_g},
\label{fem}
\end{equation}
where $\sigma$ is conductivity, $n$ is the electron concentration, $q$ is electron charge, $C'_g$ is the gate capacitance per unit length and $L$ is the channel length. Equation~\ref{fem} only strictly holds at peak mobility, where $\frac{d\mu_{\text{fe}}}{dn}=0$. The effective mobility is defined as 
\begin{equation}
\mu_{\text{eff}}  = \frac{LG}{C'_g(V_g-V_t)},
\label{effm}
\end{equation}
where $V_t$ is the pinchoff threshold voltage defined previously, and the expression only holds for $V_{sd}<<V_g-V_t$. The two mobility measures are compared in figure~\ref{fig3}(a) for device 2 at 40 K. The effective mobility is typically a smoother function of $V_g$, and $\mu_{\text{eff}}\geq\mu_{\text{fe}}$ for all of our data. Two regimes can be clearly seen in $\mu_{\text{eff}}$: the slope $|\frac{d\mu_{\text{eff}}}{dV_g}|$ is larger from $V_g= -0.25V$ to $V_g= +0.25V$ than at more positive gate voltages. In figure~\ref{fig3}(b) we show the effective mobility versus temperature for device 2 at different values of gate voltage relative to the position of peak effective mobility ($V_{peak}$). The data shown are for $V_g=V_{peak}+\delta$, where the top curve (black dots) is for $\delta=0$, and the lower curves (red, green, blue) are for $\delta=0.25, 0.5, 1.0$ V, respectively. The temperature dependence is most pronounced at peak mobility, but follows a similar trend for points on the high slope region of the effective mobility curve. At large positive gate voltages relative to $V_{peak}$, the mobility shows little to no dependence on temperature. We ascribe the gate dependence of effective mobility, which for all devices is qualitatively similar to that shown in figure~\ref{fig3}(a), mainly to a surface accumulation layer of electrons that forms as the gate is made more positive. This accumulation layer will act to screen the conduction electrons in the core of the nanowire, effectively reducing the gate capacitance to the core electrons and producing a smaller observed mobility, since we do not take this screening into account in equation~\ref{effm}. As the electron density in the accumulation layer becomes larger, it also dominates the device conductance and has a lower intrinsic mobility due to its proximity to the surface. The peak mobility, however, occurs close to pinchoff where the accumulation layer should be absent or negligible. At peak mobility, the nanowire surface potential is close to the flat-band condition, and we would also expect little or no interface capacitance \cite{dayeh2007b} as long as the gate sweep is sufficiently slow. Hence, the peak mobility should be a good approximation to the intrinsic mobility of the conduction electrons in the bulk of the nanowire, so that is the quantity we focus on in the remainder of the paper. \\
\indent A possible source of systematic error in mobility is shielding due to Ohmic contacts \cite{Pitanti2012}, which can become large for short channel lengths. For our shortest channel length of 770 nm (device 3), the calculated mobility could be overestimated by up to a factor of two in the worst case. The shielding error should be negligible for device 1. This type of error is independent of temperature, and therefore does not affect the qualitative behaviour of mobility. Another concern is the dependence of the measured mobility on the bias voltage. We observe no difference, within statistical error, between mobilities measured at 1 mV and 10 mV bias. Recently, challenging Hall effect measurements were carried out \cite{Storm2012,Blomers2012} on InAs nanowires showing that immobile interface charge accounts for an appreciable fraction of the the total gate-induced charge, meaning that field-effect measurements tend to underestimate the true mobility. We argue that this mechanism would most strongly affect the mobility estimates in the device ``on" state rather than at peak mobility where the surface potential is nearly flat. Therefore we expect that the qualitative temperature dependence we measure reflects intrinsic behaviour and is not an artifact of interface capacitance effects. Temperature and gate-dependent Hall measurements on our (relatively smaller diameter) nanowires are desirable to confirm this, but are beyond the scope of this paper.\\
\indent Devices 1 and 3 show qualitatively similar behaviour to device 2, as shown in figure~\ref{fig4}(a). The maximum in mobility at around $T=50$ K is consistent with previous reports \cite{ford2009, Dhara2011}. At a given temperature, the mobility increases with nanowire diameter, as was also reported previously \cite{ford2009}. This is consistent with the mobility being dominated by surface charge scattering, as the overlap of the carrier distribution with the scattering potential becomes much stronger at smaller diameters \cite{Das2005}. Note, however, that we have not examined enough devices to make firm conclusions about diameter dependence on statistical grounds. Motivated by the hypothesis that surface scattering dominates the mobility, the data in figure~\ref{fig4} are fit to empirical function of the form $\mu(T) \propto T^x N(T)^{-y}$, where $N(T)$ is the number of surface scatterers. This function does not result from an analytical solution of the surface scattering problem, which is in general too difficult to solve without resorting to numerics \cite{Das2005}. Rather, this function provides a good model for our data and is based on the the following reasoning. For a fixed number of scatterers, the average mobility increases with temperature as $T^x$, where $x\sim 1$, since the carrier concentration increases with temperature leading to an increase in the Fermi velocity, which reduces the scattering probability \cite{Das2005, nag1980}. On the other hand, an increase in the number of scatterers decreases mobility. In the limit of a low density of scatterers and a high probability of scattering per defect, scattering events can be treated as uncorrelated, and $\mu \propto N^{-1}$ (or equivalently, the scattering rate is proportional to the number of scatterers). However, for scattering from positively charged surface states, there is a high density of scatterers with a low probability of scattering per defect, leading to correlated scattering \cite{Evans2005} (see section below). Here, the electron wavefunction remains coherent while interacting with multiple surface charges simultaneously, which leads roughly to $\mu \propto N^{-2}$, since the scattering matrix element is roughly proportional to $N$, so the transition rate is proportional to $N^2$. We model $N(T)$ based on the thermal activation of surface donors: $N(T) \propto (1+Be^{-E_a/kT})$, where $B$ and $E_a$ are free parameters, similar to the expression used in figure~\ref{fig2} to model the pinchoff threshold voltages. \\
\indent The data in figure~\ref{fig4} are fit to $\mu = A T^x (1+Be^{-E_a/kT})^{-2}$. For $D=(71, 50, 35)$ nm, the fit parameters (excluding scaling factor $A$) are the following: $x=(1.0, 1.25, 0.67)$, $B=(13.4, 14.6, 3.0)$, and $E_a=(17.2, 15.1, 8.0)$ meV. We note that the data can be fit equally well to a functional form $\mu \propto N^{-1}$, albeit with different fit parameters, but we chose the $N^{-2}$ form for consistency with the numerical modelling results in the next section. The $E_a$ values suggest thermal ionization of the surface donor states with activation energies in the range $8-20$ meV, consistent with the range of $E_a$ values obtained from fitting $V_t$ in figure~\ref{fig2}. The smaller value of $B$ for the 35 nm diameter nanowire is consistent with the weaker temperature dependence of its pinchoff threshold voltage in figure~\ref{fig2}(f), indicating a smaller number of thermally activated donor states relative to the larger diameter nanowires. Figure~\ref{fig4}(b) compares the data for device 1 before and after an ammonium sulfide etching and passivation treatment was applied to the FET channel. The best fit parameters in the latter case are $x=0.62$, $B=4.5$, $E_a=20.2$ meV. After the chemical treatment, the turnover in mobility broadens and shifts to higher temperatures. This is accompanied by a much weaker change in the pinchoff threshold voltage with temperature, shown in figure~\ref{fig2}(d). The smaller value of fit parameter $B$ after chemical treatment is consistent with the weaker temperature dependence of pinchoff threshold voltage after treatment. We note here that the detailed condition of the nanowire surface post-treatment is not known, and it is likely that the native oxide partially or fully regrew before or during the post-treatment transport measurements. The data are presented only to show that the nanowire transport properties are significantly altered by chemical removal of the oxide followed by unknown surface chemical processes; these processes evidently incur some change in the nature or density of surface states. The overall reduction in mobility is consistent with previous observations of low mobility in nanowires exposed to wet etching conditions \cite{Storm2012, Dhara2011}, which could be due to changes in surface states, increased surface roughness, or a combination of the two. \\
\section{Numerical modelling}
\begin{figure*}[!t]
\includegraphics[width= 15cm]{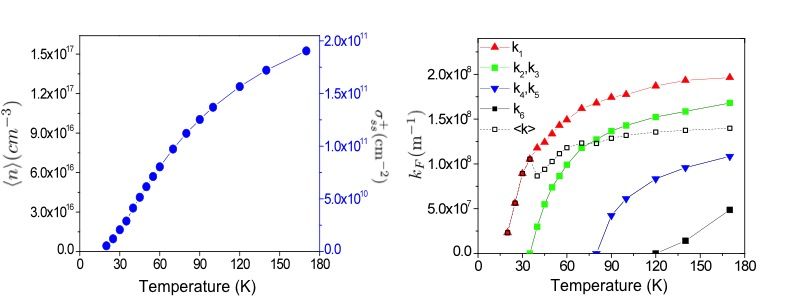} 
\caption{(a) The values of surface donor density, $\sigma^{+}_{ss}(T)$, used as inputs for the numerical simulation of a 50 nm diameter nanowire are shown on the right vertical axis. The functional form, described in the text, models a simple thermal activation of donors. The resulting average conduction electron density $\langle n \rangle$ is shown on the left axis. The $\sigma^{+}_{ss}(T)$ values were chosen to produce $\langle n \rangle$ at $V_g=0$ similar in magnitude to the values observed experimentally for device 2 at peak mobility. (b) Fermi wavenumbers $k_1,..,k_6$ of the first six radial subbands calculated from the Schrodinger-Poisson solutions for inputs $\sigma^{+}_{ss}(T)$. $\langle k \rangle$ is the average value over thermal occupation, and is proportional to the average electron velocity.} \label{fig5}
\end{figure*}

\begin{figure*}[!t]
\includegraphics[width= 15cm]{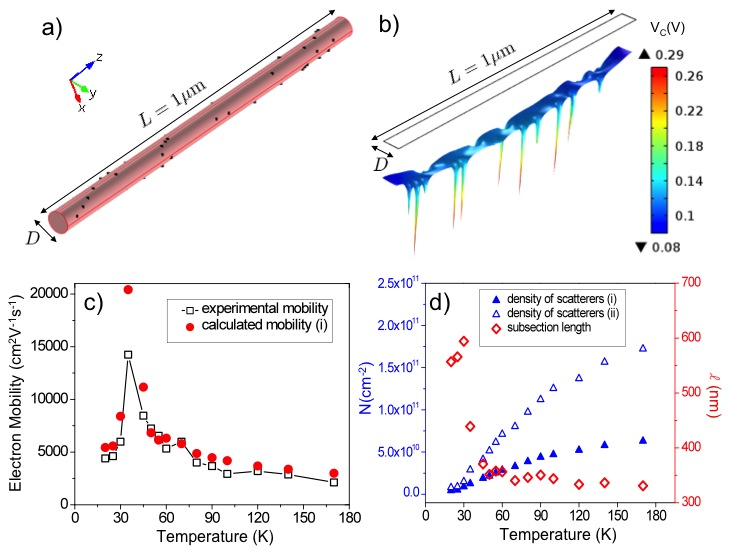} 
\caption{(a) Geometry used for calculating scattering from a random distribution of surface charges for a nanowire of total length $L =$ 1 $\mu$m and diameter $D = $50 nm. The total scattering rate is obtained by calculating the scattering matrix elements over the entire nanowire in method (i), or by calculating the matrix elements over a subsection of length $l$ and incoherently adding the rates from all $L/l$ sections in method (ii). (b) Poisson potential $V_C$ corresponding to the surface charge distribution in (a), projected onto a plane along the axis of the nanowire. (c) Comparison of the experimental mobilities (device 2) and the mobilities calculated using method (i) (the results using method (ii) are nearly identical). (d) The densities of surface charges $N(T)$ that produce the calculated mobilities in (c) for both methods. The subsection lengths $l$ used in method (ii), loosely identified with mean free path, are shown on the right axis.} \label{fig6}
\end{figure*}

\indent We carried out numerical modelling of the nanowire transistor to test whether scattering from charged surface states can account for the magnitude and temperature dependence of the experimental mobilities. The nanowire transistor was simulated using a finite-element method implemented in the COMSOL\textsuperscript{\textregistered} multiphysics package. The  model consisted of a 1 $\mu$m long, 50 nm diameter nanowire atop a 175 nm thick SiO$_2$ layer with underlying backgate. The layer above the SiO$_2$ that embeds the nanowire is vacuum, with $\epsilon_r=1$, and we take $\epsilon_{r}=15.15$ for the InAs nanowire. In consideration of the low effective mass of electrons in InAs, we used a self-consistent Poisson-Schrodinger solver \cite{Datta2005} to calculate the electrostatic potential and charge distribution in the nanowire so that quantum confinement is properly taken into account. The model assumes that the conduction electron concentration at zero gate voltage is due to a surface density of positively charged donor states, $\sigma^{+}_{ss}\sim 10^{11}-10^{12}$ cm$^{-2}$, an input parameter that is allowed to vary with temperature.\\
\indent Consider a Cartesian coordinate system with $z$ aligned with the nanowire axis and radial coordinates $(x,y)$. The potential $V(x,y,z)$ that is a solution to the Poisson equation is nearly independent of axial coordinate $z$, so we solve the Schrodinger equation in a two-dimensional cross-section of the nanowire to obtain the radial eigenstates $\psi_i(x,y)$. The electron density as a function of the radial coordinates $n(x,y)$ is calculated from these solutions as
\begin{equation}
n(x,y)=\sum_i n_{i}(x,y) = \sum_i|\psi_i(x,y)|^2\int_{E_i}^{\infty}f(E)g(E-E_i)dE
\label{nxy}
\end{equation}
where $g(E-E_i)=\frac{L}{\pi\hbar} \sqrt{\frac{2m^*}{E-E_i}}$ is the one-dimensional (1D) density of states, $f(E)$ is the Fermi-Dirac distribution, and $E_i$ and $\psi_i(x,y)$ are the energy and wavefunction of the $i^{th}$ eigenstate, respectively. The average electron concentration is obtained by integrating over the radial coordinates and dividing by the volume $\pi R^2 L$. A change of variables $E \rightarrow (E-E_i)/kT$ leads to a compact form: 
\begin{equation}
\langle n \rangle = \frac{\sqrt{2m^*k_BT}}{\pi^2 \hbar R^2}\sum_{i} F_{-1/2}\left(\frac{E_F-E_i}{k_BT}\right),
\label{density}
\end{equation}
where $E_F$ is the quasi-Fermi level, $F_{-1/2}$ is the Fermi-Dirac integral of order $-1/2$, and $m^*$ is $0.023$ times the electron mass. The Fermi energy $E_F$ is determined by the net conduction electron concentration at zero gate voltage. Figure~\ref{fig5}(a) shows the values of $\sigma^{+}_{ss}(T)$ used in the simulations, and the resulting average conduction electron density $\langle n \rangle$ versus temperature. We chose a function $\sigma^{+}_{ss}(T) = \sigma_0+\sigma_1e^{-E_a/kT}$ to model the thermal activation of surface donor states, where $\sigma_0 = 1.7\times10^9$ cm$^{-2}$, $\sigma_1 = 9.8\times10^{10}$ cm$^{-2}$ and $E_a = 6.7$ meV for the curve in figure~\ref{fig5}(a). These values were chosen so that the simulated electron density at zero gate voltage would roughly match the experimentally measured carrier density of device 2 at peak mobility. Note that peak mobility occurred at negative gate voltages in the real device, so the actual densities of surface donor states are likely larger than the values used in simulation. The reason for carrying out the simulations at zero gate voltage was to model the behaviour for a radially symmetric wavefunction, unperturbed by the presence of a nonzero gate voltage, for simplicity. \\
\indent Mobility is calculated using a multi-subband momentum relaxation time approximation \cite{Ferry1997}. We define three-dimensional eigenstates $|m,k\rangle = \psi_m(x,y) e^{i k z}/\sqrt{L}$, where $m$ is the radial subband index and $k$ is the axial wavenumber. The transition probability $T_{k,k'}^{mn}$ between the states $|m,k\rangle$, $|n,k'\rangle$ are calculated using Fermi's golden rule:
\begin {equation}
T^{mn}_{k,k'}=\frac{2\pi}{\hbar}|M^{mn}_{k,k'}|^2\delta(E_k-E_{k'})
\end{equation}
where $M^{mn}_{k,k'}$ is the scattering matrix element $\langle k,m|V_C|k',n\rangle$ resulting from the Coulomb interaction potential $V_C$ of charged surface impurities. In our numerical simulations, $V_C$ is obtained directly from the Poisson solver, and this takes into account both screening and dielectric mismatch effects \cite{Jena2007, Salfi2012}. In the absence of these effects, $V_C$ would be analytically expressed as a sum over unscreened point-charge potentials. In a cylindrical coordinate system $(r, \theta, z)$ where $r$ and $z$ are the radial and axial coordinates,
\begin{equation}
V_C=\sum_i V_{C,i} = \frac{e^2}{4\pi\epsilon_0\epsilon_r}\sum_i \left(r^2 + (D/2)^2 - rD\cos{\theta_i} + (z-z_i)^2 \right)^{-1/2}
\end{equation}
where $V_{C,i}$ is the potential due to a single impurity located at $\bold{r}_i = (D/2, \theta_i, z_i)$. With the numerically-derived $V_C$ that includes screening effects, we find that the value of $M^{mn}_{k,k'}$ for a single positively charged surface impurity is on the order of $10^{-2}$ meV or less. Its smallness is due to the vanishing of $|\psi|^2$ at the surface, the large dielectric constant for InAs, screening effects, and that the scattering potential is attractive. In this case, treating scattering from single impurities independently and incoherently adding their rates can only lead to the observed mobilities if the surface impurity charge densities are unreasonably high, $N \sim 10^{13}$ cm$^{-2}$. At such densities, the mean separation between scatterers is too small for the picture of uncorrelated scattering to be valid. On the other hand, for a $V_C$ that is the collective potential corresponding to a random distribution of many scatterers over the length of the nanowire, we are able to obtain the observed mobilities at impurity densities $N(T)\sim \sigma^{+}_{ss}(T)$ (see Figure 6). This approach justifies the empirical expression $\propto N^{-2}$ used in the previous section to fit the experimental data, since the scattering matrix element $M^{mn}$ now roughly scales with $N$, rather than being independent of $N$ in the picture of uncorrelated single-defect scattering.\\
\indent The scattering matrix element is given by 
\begin{equation}
M^{mn}_{k,k'}= \int_{0}^{D/2} \int_{0}^{2\pi} \int_{-L/2}^{L/2} r \psi_m(r,\theta) V_C \psi^*_n(r,\theta)e^{-i(k-k')z}dz d\theta dr
\label{element}
\end{equation}
where $V_C$ is the total potential corresponding to a set of impurities. The integral in equation~\ref{element} has no straightforward analytical solution, so is generally solved numerically \cite{Das2005}. The geometry for simulating correlated scattering is indicated schematically in figure~\ref{fig6}(a), and the Poisson solution $V_C$ obtained for a random impurity distribution is shown in figure~\ref{fig6}(b). The relaxation rate in subband $m$ due to scattering into subband $n$  is calculated as
\begin{equation}
1/\tau^{mn}(k)=\sum_{k'}(1-\cos{\phi}) T^{mn}_{k,k'}
\label{rate}
\end{equation}
where $\phi$ is the angle of deflection between the incoming wave vector $k$ and the outgoing wave vector $k'$. The values of $k'$ are given by energy conservation, $E_m+\hbar^2k^2/2m^{*}=E_n+\hbar^2k'^2/2m^{*}=E_F$. In a 1D geometry, only back scattering events contribute to electron relaxation rates. When the electron concentration permits the occupation of multiple subbands, the relaxation rate in the $m^{th}$ subband is obtained as $1/\tau_m(k)=\sum_{n} 1/\tau^{mn}(k)$, where $k$ is the initial momentum. At low temperatures, it is valid to only consider the relaxation time for an electron with Fermi wavenumber $k_F$. Making this approximation, we substitute the Fermi wavenumber in each subband for $k$. The average relaxation time is given by $\tau=\sum_i \tau_i n_i/n$, where $n_i$ is the population of $i^{th}$ subband, leading to an average electron mobility $\mu=e\tau/m^*$. Figure~\ref{fig5}(b) shows the Fermi wavenumbers of the first few radial subbands calculated from the Schrodinger-Poisson solutions for input donor densities $\sigma^{+}_{ss}(T)$. The first excited subband appears near 40 K, producing a dip in the average wavenumber $\langle k_F \rangle$. The sharp drop in Fermi velocity as temperature is lowered below 40 K strongly increases the ionized impurity scattering rate, which causes a drop in mobility. \\
\indent We performed the scattering calculations in two ways: (i) calculating integrals $M^{mn}_{k,k'}$ for the electron wavefunction and scattering potential over the entire length of the $L=1$ $\mu$m nanowire, and (ii) restricting the problem to a subsection of the nanowire of length $l < L$. Method (ii) is motivated by the fact that the experimentally observed mobilities suggest a mean free path $l_{mf} \sim 100-200$ nm or less \cite{dayeh2007c}, so that on average, we expect an electron traversing the nanowire to experience several uncorrelated scattering events. In the latter picture, the scattering rate $\tau^{-1}$ is calculated from the $T^{mn}_{k,k'}$ for the electron wavefunction restricted to a length $l$ comparable to the mean free path, and the scattering rate for the entire length of nanowire is $L/l$ times this rate. On the other hand, the probability for the electron to be in any one subsection is $l/L$, so these factors cancel. The only difference between the two cases is that the 1D density of states $g l$, which appears in the evaluation of equation~\ref{rate}, is proportional to the subsection length. Hence, for an electron treated quantum mechanically on a length scale $l$ (but classically on larger length scales), the density of states to scatter into is lower than if the wavefunction were spread across length $L$, increasing the calculated mobility. Therefore a factor $L/l$ larger density of scatterers is required in calculation (ii) relative to (i) in order to produce the same calculated mobility. \\
\indent The results of these calculations are shown in figure~\ref{fig6}: (d) shows the density of scatterers $N$ obtained by calculations (i) and (ii) that reproduce the experimental mobilities. In calculation (ii), a variable subsection length $l$ was chosen such that $N(T)\approx \sigma^{+}_{ss}(T)$; these $l$ values are plotted on the right axis. The calculated mobilities from (ii) are shown in figure 6(c) in comparison with the experimental values. A three-fold increase of $N$ over the range 40-150 K is able to explain the observed decrease in mobility with temperature for both calculation methods. Furthermore, the density of scatterers is nearly a perfect match to the assumed ionized surface donor density for method (ii). It is reasonable to expect that the increase of $N$ with temperature results from the thermally activated ionization of surface donor states. Confinement also plays a role in this temperature dependence, since higher radial subbands contribute to a larger electron concentration near the surface, with a corresponding increased scattering rate. However, for a fixed $N$, this confinement effect is too small to cause a negative slope of the mobility-versus-temperature. We find that interband scattering plays a very limited role, giving at most a correction of order $10\%$ to the scattering rates. As expected, the positive slope of mobility below 40 K follows the behaviour of the average Fermi velocity (figure~\ref{fig5}(b)) over the same temperature range, where only the lowest radial subband is occupied. Overall, the simulation results confirm that scattering from charged surface states at densities typical of InAs can explain the magnitude and temperature dependence of the experimental mobilities. \\
\begin{figure*}[!t]
\includegraphics[width= 14cm]{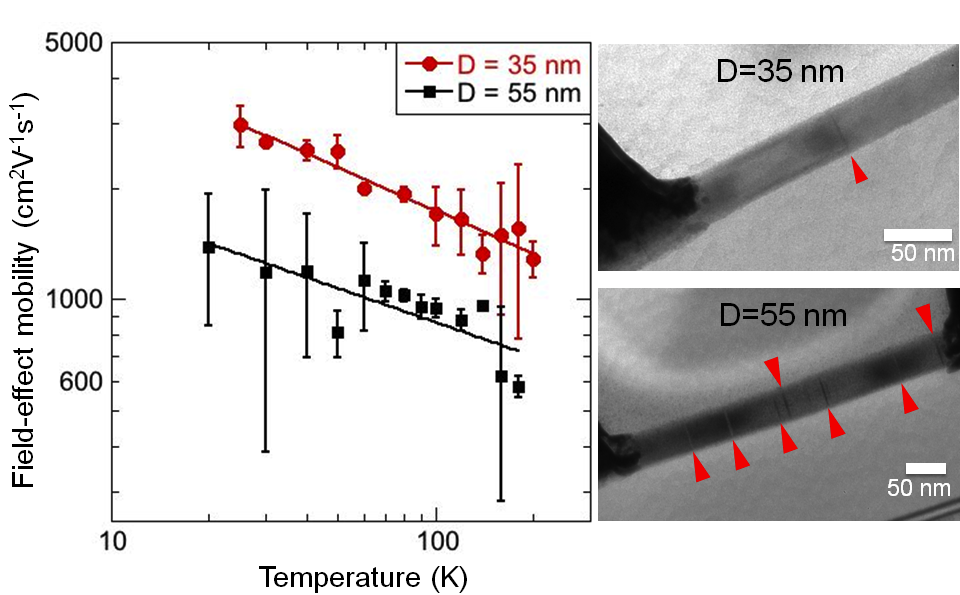} 
\caption{Stacking fault density and reduced mobility. Peak field-effect mobilities (left) and post-measurement TEM images (right) for device 3 ($D = $35 nm) and a low-mobility $D = $55 nm nanowire FET device. Stacking faults are indicated by the red arrows; at least $7$ faults can be seen in the $D = $55 nm nanowire, compared to only one visible fault in the $D = $35 nm nanowire. The nanowires are imaged along the [$2 \bar{1} \bar{1} 0$] zone axis so that all planar defects will be visible. The solid lines show power law fits to $T^{-0.4}$ and $T^{-0.3}$ for the 35 nm and 55 nm devices, respectively. No faults were observed along the entire channel for device 1 (average diameter 71 nm).} \label{fig7}
\end{figure*}
\section{Structural defects and mobility}
\label{faults}
\indent Finally, we studied the relationship between structure and mobility by performing post-measurement TEM on selected devices; this was motivated by the observation that a fraction of devices displayed significantly lower mobilities than were typical for a given nanowire diameter. A Focused Ion Beam (FIB) was used to remove devices from the substrate, after which they were placed on a holey carbon TEM grid for inspection. Indeed, it was observed that a $55$ nm diameter nanowire with low mobility $\sim 1,000$ cm$^2$/Vs had a high linear density of stacking faults, at least $\sim$($70$ nm)$^{-1}$ as shown in figure \ref{fig7}. In contrast, the highest mobility device we measured, device 1, had no visible faults along the entire channel length. Device 3 ($D=35$ nm) was found to have only one visible fault as shown in figure \ref{fig7}, and better mobility than the $D=55$ nm device, despite having a smaller diameter. The magnitude and temperature dependence of mobility appear to be greatly reduced in the $D=55$ nm device due to the high density of stacking faults. Wurtzite InAs has a $\sim 20\%$ larger bandgap than zincblende InAs \cite{bao2009}, so that for electrons, stacking faults correspond to potential wells that may be as deep as $\sim 70$ meV. Since these are planar defects, the reflection coefficient for an incoming plane wave can be a sizable fraction of unity. On the other hand, we cannot obtain theoretical mobilities as low as $\sim 1,000$ cm$^2$/Vs from a simple 1D model of square well potentials at the linear defect density observed here. It is possible that the longer zincblende sections may contain bound states that trap electrons \cite{Wallentin2012}, leading to Coulomb scattering. Gap states that trap charges locally can arise at dislocations \cite{Ebert2001}, however, there are no mechanisms within the VLS growth method through which dislocations could form for the bare (111) oriented InAs nanowires studied here. A stacking fault is simply a rotation of the tetrahedral coordination for one monolayer, which leaves the lattice four-fold covalently bonded and free of distortion. Further investigation is required to clarify the origin of the surprisingly low mobilities seen here. Importantly, the low fault densities observed in devices 1 and 3, together with the characteristic mobility temperature dependence in figure 4, rules out the possibility of stacking faults being responsible for the turnover in mobility below 50 K. \\
\section{Discussion}
\indent While the data and modelling in sections 3 and 4 are consistent with a dominant role of positively charged surface states as scatterers, it is also possible that negatively charged impurities, such as native oxide charge traps \cite{Salfi2011}, might play a role. Negative charges produce stronger scattering potentials \cite{Salfi2012}, so that a relatively small number of impurities could limit the electron mobility. On the other hand, we observe that the pinchoff threshold voltage shifts to more positive values as temperature is reduced, but more positive gate voltages should lead to \emph{higher} occupation of negative traps. Furthermore, if oxide charge traps limited mobility, then we would expect much higher mobilities in epitaxial core-shell nanowires where the oxide surface is $10-20$ nm away from the core. Somewhat higher mobilities were observed in those nanowires \cite{tilburg2010}, but only by a factor $\sim 1.4$ compared to the best results with unpassivated nanowires reported elsewhere \cite{ford2009} and in the present work. We suspect this improvement in mobility is due to passivation of surface states rather than moving oxide charge traps further away from  the channel. Further experiments on chemically and epitaxially passivated nanowires may test this hypothesis. A related concern is the possibility of scattering due to electrostatic fields from trapped charges in the underlying SiO$_2$ substrate. This cannot be firmly ruled out from the present data, but could be addressed by future experiments on suspended FET devices. Surface roughness scattering might also limit the mobility, and it is not clear from the literature what temperature dependence to expect, although there is some indication it should be weak \cite{nag1980}. From high-resolution TEM we estimate a typical roughness less than 2-3 monolayers for these nanowires. We expect surface roughness to play a more significant role in the low mobility of the accumulation layer than in limiting the mobility of the bulk conduction electrons. Especially at low temperature and close to pinchoff, the electron distribution is predominantly in the center of the nanowire, with vanishing probability at the surface. Hence, Coulomb scattering should dominate over neutral defects like surface roughness if the density of surface charges is sufficiently high ($\sim10^{11} - 10^{12}$ cm$^{-2}$). At low temperatures we must also consider the Coulomb interaction between electrons that form `puddles' in a disordered potential, i.e. charging effects. This might provide an alternate explanation for the observed mobility drop below 50 K. However, we have recently observed an opposite trend in InAs-In$_{0.8}$Al$_{0.2}$As core-shell nanowires \cite{Holloway2013a}, in which the mobility continues to increase as temperature is lowered, despite the fact that strong, qualitatively similar Coulomb oscillations appear below $\sim 10$ K in both types of nanowire. We ascribe the difference in mobility behaviour to a reduction of InAs surface states by the epitaxial passivation. \\
\section{Conclusion}
\indent In conclusion, our data and numerical simulations support the hypothesis that ionized impurity scattering by charged surface states dominates the peak electron mobility in low defect density InAs nanowires across a wide range of temperatures. Transport measurements show a ubiquitous turnover in the temperature-dependent mobility below $\sim$50 K. The behaviour above 50 K can be explained by a thermally activated increase in the number of ionized scatterers. These results on pure InAs nanowires provide a benchmark to compare with the transport behaviour of heteroepitaxial core-shell nanowires or nanowires with stable chemical passivation. Additionally, post-transport TEM measurements show that a high stacking fault density, observed in a small fraction of these nanowires, leads to sharply reduced mobilities and a weaker temperature dependence. \\ 
\ack{We thank the Canadian Centre for Electron Microscopy, Julia Huang and Fred Pearson for technical help with FIB and TEM; the Centre for Emerging Device Technologies and Sharam Tavakoli for assistance with MBE; Om Patange and David G. Cory for use of AFM; the QNC Nanofabrication facility and its supporting agencies. This work benefitted from discussions with Milad Khoshnegar, Daryoush Shiri and Mohammad Ansari. This research was supported by NSERC, the Ontario Research Fund, the Canada Foundation for Innovation and Industry Canada. G. W. H. acknowledges a WIN Fellowship.}

\section*{References}
\bibliographystyle{iopart-num}
\bibliography{nanowire.bib}

\end{document}